\documentclass[universe,article,accept,pdftex,oneauthor]{Definitions/mdpi}
\firstpage{1}
\pubvolume{1}
\issuenum{1}
\articlenumber{0}
\pubyear{2025}
\copyrightyear{2025}
\externaleditor{Firstname Lastname}
\datereceived{ }
\daterevised{ } 
\dateaccepted{ }
\datepublished{ }
\hreflink{https://doi.org/} 
\usepackage{amsfonts,amssymb,amsmath,graphicx,multirow,dsfont}
\def\dac{\displaystyle\frac}

\def\[{\left[}
\def\]{\right]}
\def\({\left(}
\def\){\right)}

\newcommand{\diag}{\mathop{\rm diag}\nolimits}

\Title{On the Properties of the Power-Law Cosmological Solutions in Lovelock Gravity}

\TitleCitation{On the Properties of the Power-Law Cosmological Solutions in Lovelock Gravity}

\Author{Sergey Pavluchenko
 \orcidA{}}

\AuthorNames{Sergey Pavluchenko}

\AuthorCitation{Pavluchenko, S.}

\address[1]{University of Illinois Urbana-Champaign,
 Champaign, IL 61820, USA; sergey.pavluchenko@gmail.com}

\abstract{In this paper we study the properties of Kasner cosmological solutions in Lovelock gravity. Recent progress in the investigation of flat cosmological models
in Lovelock gravity unveiled the fact that in quadratic (Gauss--Bonnet) and cubic Lovelock gravities, the higher-order power-law solutions could play the role of both future and past asymptotes, and under some conditions, there could exist a smooth transition between them. So it is natural to question if this feature is unique to Gauss--Bonnet and cubic Lovelock gravities, or if it is a general feature of Lovelock gravity. Our analysis suggests that starting from quartic and in all higher-order Lovelock gravities, the high-order Kasner solution cannot play the role of a past asymptote, not only preventing the abovementioned transition from happening, but also potentially hindering the possibility of reaching viable compactification.}



\begin{document}

\section{Introduction}

The idea that our space has more than three dimensions is now new---first, it was introduced by  G.~Nordstr\"om~\cite{Nord} as part of his gravity model, where the extra dimension incorporated Maxwell's electromagnetism. 
Soon after, however, Nordstrom's gravity was cast aside based on solar eclipse observations in 1919, where light bending due to the Sun's gravity was detected to be in agreement with Einstein's General Relativity (GR). The idea of extra dimensions, however, survived, and very soon, a similar theory based on GR was proposed by T.~Kaluza~\cite{Kaluza}. Soon after that O. Klein supplied the model with a clear quantum mechanical interpretation~\cite{Klein1, Klein2}, and the resulting theory, called Kaluza--Klein after its two main contributors, was established. It is worth noting that, at the time of formulation, the Kaluza--Klein theory united all interactions known at that time. Presently, it is M/string theory that is considered promising to unify all interactions.

Typically, the Lagrangian of the gravitational counterpart of M/string theories contains curvature-squared corrections---$R^2$ and $R_{\mu\nu} R^{\mu\nu}$~\cite{SchSch}, as well as $R_{\mu\nu\lambda\rho} R^{\mu\nu\lambda\rho}$~\cite{Candelas}---and~in order to result in a ghost-free gravitational interaction, these terms should enter as a combination known as the Gauss--Bonnet (GB) term
 \cite{Zw}:

$$
L_{GB} = R_{\mu\nu\lambda\rho} R^{\mu\nu\lambda\rho} - 4 R_{\mu\nu} R^{\mu\nu} + R^2.
$$

Apparently,
 this term is an Euler topological invariant in (3+1
)-dimensional spacetime, but in higher dimensions, it gives a non-trivial contribution to the equations of motion. If we take higher-than-quadratic curvature terms into consideration, the low-energy limit of the united theory could be considered a sum of different powers of curvature~\cite{Zumino}. 
Probably the best-known and well-developed gravity models of this kind are those found in Lovelock gravity~\cite{Lovelock}.

Lovelock gravity is a generalization of GR in the following sense: the Einstein tensor is the only symmetric and conserved second-rank tensor, which depends on a metric and its first and second derivatives (with a linear dependence on the second derivatives); if we drop the linearity requirement for the second derivatives, we would naturally derive the Lovelock tensor~\cite{Lovelock}. This way, the zeroth-order Lovelock contribution is a constant (associated with the cosmological constant), the first order is GR (linear on curvature), the second order is the GB term (quadratic on curvature), and so on. On top of that, each Lovelock contribution is an Euler topological invariant up to a certain number of space-time dimensions: the GB term starts to manifest itself only starting from (4+1)-dimensional spacetime, the cubic Lovelock term---starting from (6+1)-dimensional spacetime, and so on.

As Lovelock gravity is formulated in a spatial dimension number higher than three, the natural question of dimension reduction arises. Indeed, all our senses and experimental data suggest that we live in three spatial dimensions, and yet, we consider a multidimensional cosmological model. The most accepted answer to this question is that the extra spatial dimensions are compactified on a scale much smaller than the detection threshold. When further investigating the compactification issue, one approach is the consideration of ``spontaneous compactification''~\cite{spont1, spont2, spont3}, and another is ``dynamical compactification''. The difference between these two approaches is, roughly speaking, that, in the former, you put compactification ``by hand'' and study the stability and/or properties of the resulting configuration, while in the latter, the system naturally evolves into a compactified configuration. Of course, the second approach provides a more natural outcome, but usually, it is a more complicated approach; notable results of this approach include those in ~\cite{dyn1, dyn2, dyn3, dyn4}. It is also worth noticing that the spatial curvature of extra dimensions, as well as initial anisotropies, could affect reaching compactification~\cite{CGP1, CGP2, CGPT, PT2017, our20, CP21, fr1, fr3, fr4, fr2}. Finally, there are studies of Friedmann-like cosmological models in Einstein--Gauss--Bonnet (EGB)~\cite{EGBF1} and Lovelock gravities~\cite{LF1} (i.e.,~without compactification).
Apart from the cosmological considerations, there have been a great deal of investigations dedicated to gravitational aspects of Lovelock gravity: studies of black holes in GB~\cite{BHGB1, BHGB2, BHGB3, BHGB5, BHGB6, BHGB7, BHGB8, BHGB9, BHGB10} and Lovelock~\cite{BHL1, BHL1.5, BHL2, BHL3, BHL4, BHL5, BHL6} gravities, some aspects related to gravitational collapse among these theories~\cite{coll1, coll2, coll3, coll4}, properties of spherically symmetric solutions~\cite{addn_8}, nonlinearity-induced shock formation~\cite{shock}, and many~others.

On a spatially flat background, the most natural future asymptote that admits spontaneous compactification is probably an exponential solution with expanding three dimensions and contracting extra dimensions. These solutions have been studied in Lovelock cosmology for ages~\cite{Is86, exp_diff1, exp_diff2, exp_diff4, exp_diff5,  exp_diff7, exp_diff8, exp_diff9, exp_diff10, exp_diff11, exp_diff12, exp_diff13}. Notable recent results include a separate description of exponential solutions with variable~\cite{CPT1} and constant~\cite{CST2} volume; we could also refer to~\cite{PT} for a discussion about the link between the existence of power-law and exponential solutions and for a discussion about the physical manifestations of different branches of solutions within these theories. Let us also mention the full description of the general scheme for finding all possible exponential solutions in arbitrary dimensions and with arbitrary Lovelock contributions taken into account~\cite{CPT3}. Apparently, not all exponential solutions listed in~\cite{CPT3} are stable~\cite{my15}; a more general approach to the stability of exponential solutions in EGB gravity could be found in~\cite{iv16}, while in~\cite{stab_add1, stab_add2, stab_add3, stab_add4}, some particular cases are described more closely.

On the other hand, power-law solutions in Lovelock gravity~\cite{spont2, Deruelle1, mpla09, iv1, iv2, grg10, prd09, prd10} typically serve as past asymptotes. However, in EGB and cubic Lovelock gravities, the Kasner solution (which is a power-law solution) could serve as a future asymptote as well (see~\cite{my16a, my16b, my17a, cubL1, cubL2,part_rev, my24} for review). So the natural question arises---is this sort of behavior limited only to EGB and cubic Lovelock cases, or is it typical for other Lovelock cosmologies as well? This is the question we are going to investigate in the current study.

It is worth mentioning that some particular cases of Kasner solutions in Lovelock gravity were studied in~\cite{CDM2015}, while here, we provide a comprehensive and generic approach that allows us to draw strong conclusions.

The manuscript is structured as follows: first, we define our model, providing the equations we will be using, as well as their derivation details; afterwards, we consider several individual cases to see their dynamics, followed by the general case, which would generalize the findings from these individual cases. Finally, we discuss the results and draw the conclusions.

\section{The~Model}

As we are going to study the power-law solutions, it is natural to consider a spatially flat background; on the other hand, our space is anisotropic, so we
start from a Bianchi-I-type metric
\begin{equation}
\begin{array}{l}
ds^2 = \diag(-1, a_1^2(t), a_2^2(t), \dots, a_{\tilde{D}}^2(t)),
\end{array} \label{metric1}
\end{equation}

\noindent where $a_i(t)$ is scale factor corresponding to the $i$th spatial dimension and $\tilde{D}$ is the total number of spatial dimensions. To obtain equations of motion, we consider Lovelock invariants~\cite{Lovelock}

$$
L_n = \frac{1}{2^n}\delta^{i_1 i_2 \dots i_{2n}}_{j_1 j_2 \dots j_{2n}} R^{j_1 j_2}_{i_1 i_2} \dots R^{j_{2n-1} j_{2n}}_{i_{2n-1} i_{2n}},
$$

\noindent where $\delta^{i_1 i_2 \dots i_{2n}}_{j_1 j_2 \dots j_{2n}}$ is the generalized Kronecker delta of the order $2n$. Then the Lagrangian density takes the form
\begin{equation}
\begin{array}{l}
{\cal L}= \sqrt{-g} \sum_n \alpha_n L_n,
\end{array} \label{lagr_gen}
\end{equation}

\noindent where $g$ is the determinant of the metric tensor, $\alpha_n$ is the coupling constant, and summation over all $n$ in consideration is assumed. The~rest of the derivation is quite straightforward (see, e.g.,~\cite{prd09}), and the resulting equations could be rewritten in terms of Hubble parameters ($H_i = \dot a_i/a_i$):
\begin{equation}
\begin{array}{l}
(2n-1) \sum\limits_{k_1 > k_2 > \dots k_{2n}} H_{k_1} H_{k_2}
\dots H_{k_{2n}} = 0
\end{array} \label{constr_gen}
\end{equation}

For the constraint equation and
\begin{equation}
\begin{array}{l}
\sum\limits^{\tilde{D}}_{\substack{m=1 \\ m\ne i}} \left[ (\dot H_m + H_m^2) \sum\limits_{\substack{\{j_1,\,j_2,\,\dots\,j_{2n-2}\}\ne\{i, m\}\\ j_1 > j_2 > \dots j_{2n-2}}}  H_{j_1} H_{j_2} \dots H_{j_{2n-2}} \right] + \\ \\  \qquad + (2n-1)\sum\limits_{\substack{\{ k_1, k_2, \dots k_{2n}\}\ne i\\ k_1 > k_2 > \dots k_{2n}}} H_{k_1} H_{k_2} \dots H_{k_{2n}} = 0
\end{array} \label{dyn_gen}
\end{equation}

\noindent for the dynamical equation corresponding to the $i$th spatial coordinate. Please note that the equations above are obtained for a vacuum case; however, since we are interested in power-law solutions in the vicinity of the initial singularity, this assumption is quite natural.

Then, to obtain the equations of motion in the final form, we make the following assumptions:

\begin{itemize}

\item We leave only the highest Lovelock contribution---indeed, since we are interested in the high-curvature regime, terms with higher orders in curvature will dominate those with lower, such that the latter could be omitted;

\item We employ the power-law {\it ansatz} $a_i(t) = t^{p_i}$, with $p_i$ being the Kasner~exponent.

\end{itemize}

Then, following~\cite{prd09}, we apply these assumptions, resulting in the equations of motion: the constraint equation~(\ref{constr_gen}) becomes
\begin{equation}
\sum\limits_{k_1 > k_2 > \dots k_{2n}} p_{k_1} p_{k_2} \dots p_{k_{2n}} = 0,
\label{constr_gen_p}
\end{equation}

\noindent where we omit the $(2n-1)/t^{2n}$ multiplier since only one
$n$ (the highest possible) is taken into account, and we deal with
a vacuum case. The~dynamical equation~(\ref{dyn_gen}) becomes
\begin{equation}
\begin{array}{l}
\sum\limits^{\tilde{D}}_{\substack{m=1 \\ m\ne i}} \left[ (p_m^2 - p_m)
\sum\limits_{\substack{\{j_1,\,j_2,\,\dots\,j_{2n-2}\}\ne\{i, m\}\\
j_1
> j_2 > \dots j_{2n-2}}}  p_{j_1} p_{j_2} \dots p_{j_{2n-2}}
\right] +
\\ \qquad + (2n-1)\sum\limits_{\substack{\{ k_1, k_2, \dots k_{2n}\}\ne i\\
k_1 > k_2 > \dots k_{2n}}} p_{k_1} p_{k_2} \dots p_{k_{2n}} = 0,
\end{array} \label{dyn_gen_p}
\end{equation}

\noindent and we again omit the $t^{2p_m-2n}$ multiplier. Then, Equations~(\ref{constr_gen_p}) and (\ref{dyn_gen_p}) could be solved to obtain
two formal solutions (\cite{prd09}):

\begin{itemize}

\item The generalized Kasner solution governed by Equation~(\ref{constr_gen_p}) and
\begin{equation}\label{sum_p}
\sum\limits_{i=1}^{\tilde{D}}p_i = (2n-1);
\end{equation}

\item The generalized Milne solution governed by Equation~(\ref{constr_gen_p}) and
\begin{equation}\label{sum_3p}
\sum\limits_{j_1 > j_2 > \dots j_{2n-1}} p_{j_1} p_{j_2} \dots
p_{j_{2n-1}} = 0.
\end{equation}

\end{itemize}

However, we have proven that the generalized Milne solution is actually just an artifact and does not correspond to any physical situation (see~\cite{PT}), such that the generalized Kasner solution is the only physical solution, and this is the solution we shall focus on.

Before moving to the main part, let us rewrite the equations in a more compact form. To do so we use the notations of symmetric polynomials (see, e.g.,~\cite{McD}):

\begin{itemize}

\item Power sum
\begin{equation}\label{power_sum}
P_k = \sum\limits_{i=1}^{\tilde{D}} p_i^k
\end{equation}

\item Elementary symmetric polynomials
\begin{equation}\label{elem_symm}
E_k = \(-1\)^k \sum\limits_{\substack{m_1 + 2m_2 + \dots + n m_n = 0 \\ m_j > 0}}\; \prod\limits_{i=1}^n  \dac{\(-P_i\)^{m_i}}{m_i ! i^{m_i}};
\end{equation}

For practical purposes it may be better to use a recursive analog of (\ref{elem_symm})---with all $P_k$ at hand from the beginning, calculate $E_k$ one after another:
\begin{equation}\label{elem_symm2}
E_k = \dac{1}{k} \sum\limits_{j=1}^n \(-1\)^{j-1} E_{n-j}P_j.
\end{equation}

\end{itemize}

Then, the~conditions for the generalized Kasner solution ((\ref{constr_gen_p}) and (\ref{sum_p})) could be rewritten as
\begin{equation}\label{genKasner}
P_1 = \(2n - 1\); \quad E_{2n} = 0.
\end{equation}

\section{Particular $n$ Cases}
Finally, let us apply the last assumption---the entire space is split into two isotropic subspaces: three-dimensional (``our Universe'') and $D$-dimensional
(``extra dimensions''; $D = \tilde{D} - 3$), with $p_H$ and $p_h$ being the corresponding Kasner exponents. Then, the power sums take form
\begin{equation}\label{gen_power}
P_k = 3 p_H^k + D p_h^k,
\end{equation}

We use $P_1 = (2n - 1)$ and (\ref{elem_symm2}) to calculate all $E_k$ up to $E_{2n}$ and then solve the resulting (\ref{genKasner}) with regard to $p_H$ and $p_h$. Let us consider several particular $n$ cases and then generalize the results to a generic $n$ case.

\subsection{$n=2$ (Gauss--Bonnet)}

\begin{figure}[H]
\includegraphics[width=0.9\textwidth, angle=0]{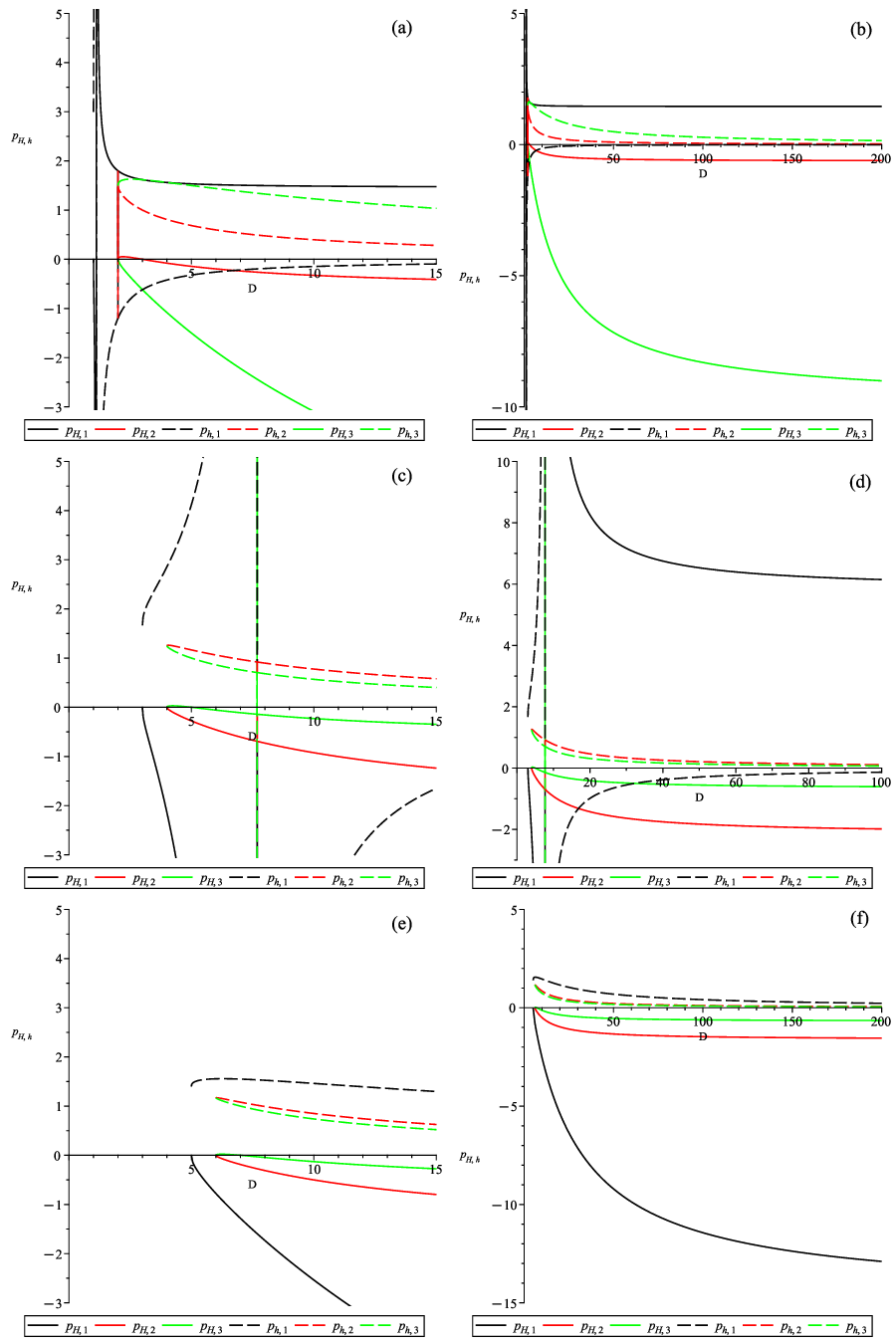}
\caption{Non-trivial
 Kasner solution for $n=2$ (panels (\textbf{a},\textbf{b})), $n=3$ (panels (\textbf{c},\textbf{d})), and $n=4$ (panels (\textbf{e},\textbf{f})) Lovelock gravities: low-$D$ details on panels (\textbf{a},\textbf{c},\textbf{e}) and high-$D$ behavior on panels (\textbf{b},\textbf{d},\textbf{f})
(see text for more~details).}\label{n2n3}
\end{figure}

In this case $P_1 = 3p_H + Dp_h = 3$, and the resulting $E_4$ reads
\begin{equation}\label{n2E4}
\begin{array}{l}
E_4^{(n=2)} = \dac{D p_h}{24} \Big( p_h^3 (D-1)(D-2)(D-3) + 12 p_H h_h^2 (D-1)(D-2) + \\ \\ + 36 p_H^2 p_h (D-1) + 24 p_H^3\Big),
\end{array}
\end{equation}

\noindent such that there is one ``static'' (with regard to $D$) solution $\{p_H, p_h\} = \{1, 0\}$ coming from the first multiplier and three ``dynamical'' (i.e., depending on $D$) solutions coming from the cubic equation. 
With the use of $P_1 = 3p_H + Dp_h = 3$, Equation~(\ref{n2E4}) could be solved and its roots plotted as a function of $D$; the resulting curves, numbered $1\div 3$, are presented in Figure~\ref{n2n3}a,b---panel (a) shows a detailed view, and panel (b) shows the large-$D$ asymptotic behavior.

One can see that there are peculiarities and low $D$ values:

\begin{itemize}

\item At $D=1$, all solutions of the cubic part of $E_4$ are trivial, leaving us with only the $\{p_H, p_h\} = \{1, 0\}$ solution;

\item At $D=2$, solutions of the cubic part of $E_4$ are $\{9/5, 0, 0\}$, such that there is only one non-trivial solution in addition to the $\{p_H, p_h\} = \{1, 0\}$ solution;

\item At $D=3$, solutions of the cubic part of $E_4$ are $\{1/2\pm\sqrt{5}/2, 0\}$ such that there are two non-trivial solutions in addition to the $\{p_H, p_h\} = \{1, 0\}$ solution, with~one of them being able to describe realistic~compactification.

\end{itemize}

Finally, starting from $D \geqslant 3$, all three solutions of the cubic part of $E_4$ are~non-trivial.

From Figure ~\ref{n2n3}(a,b) one can see that one of the $p_H$ is positive, while the two remaining are negative, and the corresponding $p_h$ have opposite signs. For three-dimensional space to expand, we need $p_H > 0$, and for an extra-dimensional subspace to contract, we need $p_h < 0$, such that, for realistic compactification, we need both $p_H > 0$ and $p_h < 0$.

Finally, let us find asymptotic $D\to\infty$ values for $p_{H, i}$. To do that we employ a trigonometric representation of the cubic roots:

\begin{itemize}

\item $\lim\limits_{D\to\infty} p_{H, 1} = 4\sqrt{3} \cos \dac{5\pi}{18} - 3 \approx 1.45336$;

\item $\lim\limits_{D\to\infty} p_{H, 2} =  4\sqrt{3} \cos \dac{7\pi}{18} - 3 \approx -0.63041$;

\item $\lim\limits_{D\to\infty} p_{H, 3} = -4\sqrt{3} \cos \dac{\pi}{18} - 3 \approx -9.82296$.

\end{itemize}

Let us note that the root appearance and numerical values coincide with the results presented in~\cite{my16a, my16b, my17a}.

\subsection{$n=3$ (Cubic Lovelock)}

In this case $P_1 = 3p_H + Dp_h = 5$, and the resulting $E_6$ reads
\begin{equation}\label{n3E6}
\begin{array}{l}
E_6^{(n=3)} = \dac{D(D-1)(D-2) p_h^3}{720} \Big( p_h^3 (D-3)(D-4)(D-5) + \\ \\ + 18 p_H h_h^2 (D-3)(D-4) +  90 p_H^2 p_h (D-3) + 120 p_H^3\Big),
\end{array}
\end{equation}
And again we have a ``static'' solution with $p_h=0$ and three ``dynamical'' solutions coming from the roots of the cubic polynomial. Similar to the Gauss--Bonnet case, we plot all three roots in Figure~\ref{n2n3}c,d---panel (c) shows a detailed view, and panel (d) shows the large-$D$ asymptotic behavior.

We again have peculiarities at low $D$:

\begin{itemize}

\item At $D=3$, all solutions of the cubic part of $E_6$ are trivial, leaving us with only the $\{p_H, p_h\} = \{5/3, 0\}$ solution;

\item At $D=4$, solutions of the cubic part of $E_6$ are $\{-15/7, 0, 0\}$, such that there is only one non-trivial solution in addition to $\{p_H, p_h\} = \{5/3, 0\}$, but~it has a contracting three-dimensional subspace, making it unrealistic;

\item At $D=5$, solutions of the cubic part of $E_6$ are $\{-285/104\pm 25\sqrt{105}/104, 0\}$, such that there are two non-trivial solutions in addition to $\{p_H, p_h\} = \{5/3, 0\}$, but~both of them have a contracting three-dimensional subspace, making them unrealistic;

\item At $D=6, 7$ there are three distinct solutions of the cubic part of $E_6$, but~all of them have a contracting three-dimensional subspace, making all of them~unrealistic.

\end{itemize}

Finally, at~$D\geqslant 8$, one of the solutions of the cubic part of $E_6$ is positive, while two others remain negative, and~this positive solution provides realistic compactification.

And again, similar to the Gauss--Bonnet case, let us find the $D\to\infty$ asymptotic values of the Kasner exponents---to do that, we again use the trigonometric representation of the cubic polynomial roots:

\begin{itemize}

\item $\lim\limits_{D\to\infty} p_{H, 1} = 1 + 2\sqrt{6} \cos \bigg( \dac{\arccos \dac{19\sqrt{6}}{54}}{3} \bigg) \approx 5.82219$;

\item $\lim\limits_{D\to\infty} p_{H, 2} = 1 - 2\sqrt{6} \sin \bigg( \dac{\pi}{6} + \dac{\arccos \dac{19\sqrt{6}}{54}}{3} \bigg) \approx -2.15934$;

\item $\lim\limits_{D\to\infty} p_{H, 3} = 1 - 2\sqrt{6} \sin \bigg( \dac{\arcsin \dac{19\sqrt{6}}{54}}{3} \bigg)  \approx -0.66284$.

\end{itemize}

Let us note that the root appearance and numerical values once again coincide with the previously reported results presented in~\cite{cubL1, cubL2}.

\subsection{$n=4$}

In this case $P_1 = 3p_H + Dp_h = 7$, and the resulting $E_8$ reads
\begin{equation}\label{n4E8}
\begin{array}{l}
E_8^{(n=4)} = \dac{D(D-1)(D-2)(D-3)(D-4) p_h^5}{40320} \Big( p_h^3 (D-5)(D-6)(D-7) +  \\ \\  + 24 p_H h_h^2 (D-5)(D-6) +  168 p_H^2 p_h (D-5) + 336 p_H^3\Big),
\end{array}
\end{equation}

And again, we have a "static" solution with $p_h=0$ and three "dynamical" solutions arising from the roots of the cubic polynomial. Similar to the previous cases, we plot all three roots in Figure~\ref{n2n3}e,f---panel (e) presents a detailed view, and panel (f) presents the large-$D$ asymptotic behavior.

Similarly to the previous cases, there are peculiarities at low $D$:

\begin{itemize}

\item At $D=5$, all solutions of the cubic part of $E_8$ are trivial, leaving us with only the $\{p_H, p_h\} = \{7/3, 0\}$ solution;

\item At $D=6$, solutions of the cubic part of $E_8$ are $\{-7/9, 0, 0\}$, such that there is only one non-trivial solution in addition to $\{p_H, p_h\} = \{7/3, 0\}$, but~it has a contracting three-dimensional subspace, making it unrealistic;

\item At $D=7$, solutions of the cubic part of $E_8$ are $\{-301/410\pm 49\sqrt{21}/410, 0\}$, such that there are two non-trivial solutions in addition to $\{p_H, p_h\} = \{7/3, 0\}$, but~both of them have a contracting three-dimensional subspace, making them~unrealistic.

\end{itemize}

Finally, at~$D\geqslant 8$, all three solutions of the cubic part of $E_8$ are distinct and negative, such that there is no realistic compactification in power-law~solutions.

At last, similar to the previous cases, let us find the $D\to\infty$ asymptotic values of the Kasner exponents---to do that, we again use the trigonometric representation of the cubic polynomial roots:

\begin{itemize}

\item $\lim\limits_{D\to\infty} p_{H, 1} = - \dac{4\sqrt{46} \cos \Bigg( \dac{\arccos \dac{307 \sqrt{46}}{2116}}{3}\Bigg)}{3} - \dac{17}{3} \approx -14.69367$;

\item $\lim\limits_{D\to\infty} p_{H, 2} = \dac{4\sqrt{46} \cos \Bigg( \dac{\arccos \dac{307 \sqrt{46}}{2116}}{3} + \dac{\pi}{3}\Bigg)}{3} - \dac{17}{3} \approx -1.62028$;

\item $\lim\limits_{D\to\infty} p_{H, 3} =\dac{4\sqrt{46} \sin \Bigg( \dac{\arccos \dac{307 \sqrt{46}}{2116}}{3} + \dac{\pi}{6}\Bigg)}{3} - \dac{17}{3} \approx -0.68605$.

\end{itemize}

Unlike previous cases, we have not thoroughly investigated cosmological models in $n=4$ Lovelock gravity, so we do not have prior results to compare with.

\subsection{Conclusions}

We have seen that cases $n=2$, $n=3$, and $n=4$ are different and that cases $n=5$ and $n=6$ (explicit results are not shown) generally follow $n=4$, such that we can assume that all $n \geqslant 4$ cases are generally the same in the sense of Kasner regimes' abundance and properties. However, to test it explicitly, we shall consider the general $n$ case and study it in the next section.

\section{General $n$ Case}

In the general case we have $P_1 = 3p_H + Dp_h = (2n - 1)$, and after some combinatorics, the resulting $E_{2n}$ takes the form
\begin{equation}\label{E2ngen}
\begin{array}{l}
E_{2n} = \dac{1}{(2n)!} \dac{D!}{(D-(2n-3))!} p_h^{2n-3} \Big( p_h^3 (D-(2n-3)) (D-(2n-2)) \times \\ \\ \times (D-(2n-1)) +  6n (D-(2n-3)) (D-(2n-2)) p_h^2 p_H + \\ \\ + 6n (2n-1) (D-(2n-3)) p_h p_H^2 + 2n (2n-1) (2n-2) p_H^3 \Big),
\end{array}
\end{equation}

And one can easily verify that substituting the corresponding $n$ into (\ref{E2ngen}) restores (\ref{n2E4}), (\ref{n3E6}), and~(\ref{n4E8}).

Similarly to the previous cases, there are peculiarities at low $D$:

\begin{itemize}

\item At $D=(2n-3)$, all solutions of the cubic part of $E_{2n}$ are trivial, leaving us with only the $\{p_H, p_h\} = \{(2n-1)/3, 0\}$ solution;

\item At $D=(2n-2)$, solutions of the cubic part of $E_{2n}$ are $\{-3(2n-1)/((2n-5)(2n+1)), 0, 0\}$, such that there is only one non-trivial solution in addition to $\{p_H, p_h\} = \{(2n-1)/3, 0\}$, but~it has a contracting three-dimensional subspace for $n\geqslant 3$, making it unrealistic;

\item At $D=(2n-1)$  solutions of the cubic part of $E_{2n}$ are

$$
\{ - \dac{(2n-1) \Big( (12n^2 - 12n - 15) \pm (2n-1) \sqrt{12n^2 - 3}   \Big)}{2 (n+1) \( 8n^3 - 28n^2 + 10n + 19 \)} , 0\}
$$

\noindent such that there are two non-trivial solutions in addition to $\{p_H, p_h\} = \{(2n-1)/3, 0\}$, but~for $n\geqslant 3$, both of them have a contracting three-dimensional subspace, making them~unrealistic.

\end{itemize}

\begin{figure}[H]
\includegraphics[width=1\textwidth, angle=0]{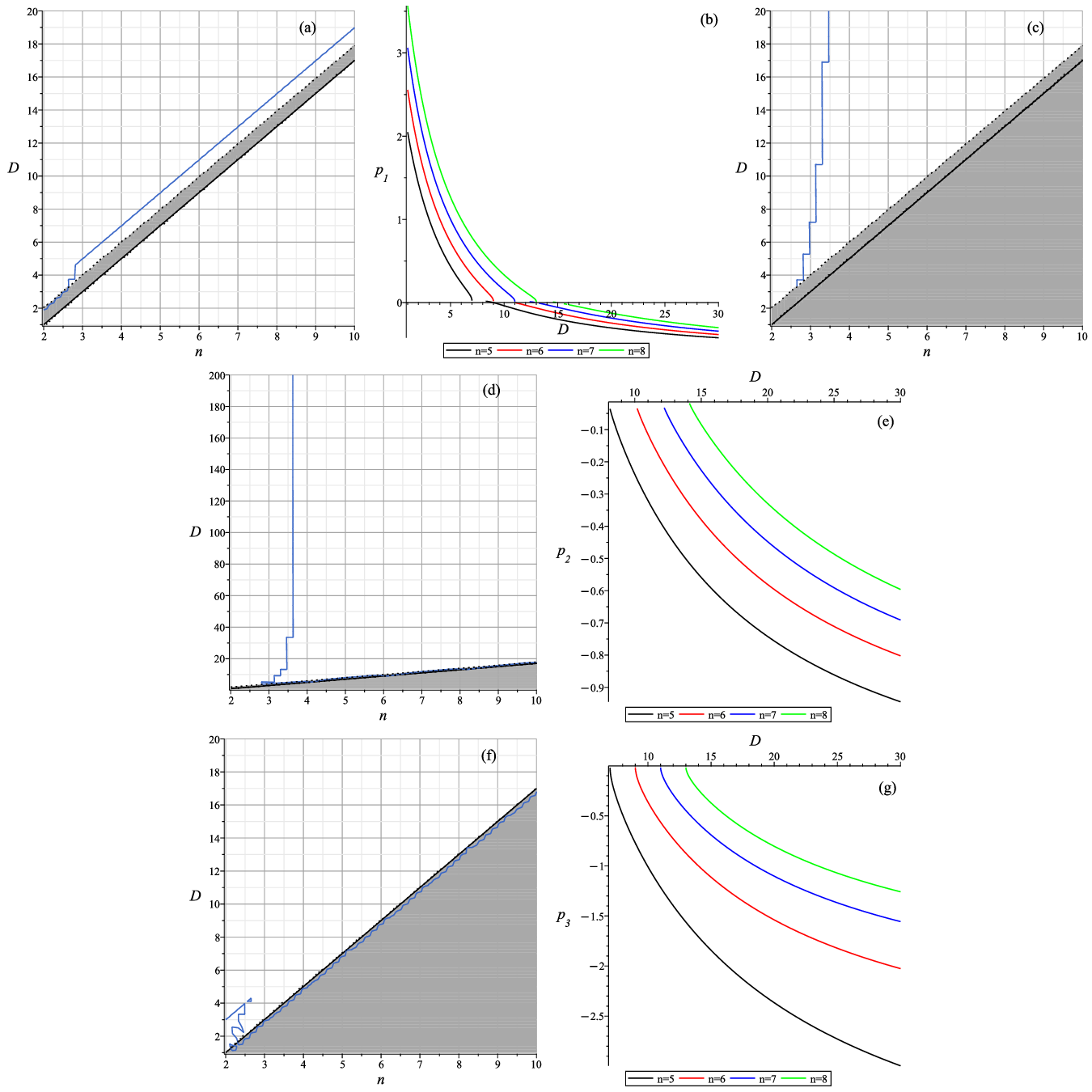}
\caption{Behavior
 of three distinct Kasner exponents: $p_1$ on panels (\textbf{a},\textbf{b}) (scan over $n$ and $D$ on panel (\textbf{a}) and several examples on panel (\textbf{b}));
 $p_2$ on panels (\textbf{c}--\textbf{e}) (scan over $n$ and $D$ on panels (\textbf{c},\textbf{d}) with large-$n$ behavior on  (\textbf{d}) and several examples on panel (\textbf{e}));
  $p_3$ on panels (\textbf{f},\textbf{g}) (scan over $n$ and $D$ on panel (\textbf{f}) and several examples on panel (\textbf{g})) (see text for more~details).}\label{genn}
\end{figure}

The discriminant of (\ref{E2ngen}) is

$$
\Delta = \dac{432 \(D - (2n-3)\)^2 (D+1)^2 (2n-1)^7 n^2 (D+2) \( d - (2n-2)\)}{D^6},
$$

\noindent such that, for $D \geqslant (2n - 1)$, it is positive and (\ref{E2ngen}) has three distinct real roots. These roots, similar to the $D\to\infty$ limits considered in the previous section, could be rewritten in a trigonometric form and plotted. Their expressions are too lengthy to be written down, but they could be easily derived from the coefficients of the cubic polynomial in  (\ref{E2ngen}), such that we plot three distinct roots of (\ref{E2ngen}), numbered as $p_{1, 2, 3}$, in~Figure~\ref{genn}. Therein, panels (a) and (b) describe the first root $p_1$; panels (c), (d), and (e) describe the second root $p_2$; and finally, panels (f) and (g) describe the third root $p_3$. Let us note that the numbering $p_{1, 2, 3}$ generally does not correspond to $p_{H, 1, 2, 3}$ used in the description of particular $n$ cases in the previous~section.

Let us investigate Figure~\ref{genn} closer. On panels (a), (c), (d), and (f), the shaded region corresponds to the case where a particular $p_i$ does not exist (rather, a real $p_i$ does not exist), the blue line on the same panels correspond to $p_i = 0$, and the black straight line depicts the limiting case $D = (2n - 3)$---minimal allowed number of extra dimensions such that the corresponding Lovelock correction is properly defined. 
In other words, $D < (2n - 3)$ is the forbidden region where the corresponding Lovelock gravity is not defined, such that we do not consider the parameter space to the bottom right of this line.

The blue line on panels (a), (c), (d), and (f) separates regions $p_i > 0$ from $p_i < 0$. On panel (a), which corresponds to $p_1$, a positive value is enclosed in a thin strip between the blue line and grey band, and there is no single integer solution, making all possible values negative. 
The situation with $p_1$ is further illustrated on panel (b), where we present the behavior of several $n$---the growing lines on the left part of the graph correspond to the forbidden $D < (2n - 3)$ area, such that, for $p_1$, all possible solutions are always non-positive.

The situation with $p_2$ is more interesting: as can be seen from panels (c) and (d), within the allowed $(n, D)$ area, there are positive values---indeed, for~$n=2$, $\forall D$, we have $p_2 > 0$, such that there always exists realistic compactification. For~$n=3$, for~low $D$ $p_2 < 0$, but starting from some $D_{cr}$, we have $p_2 > 0$. In fact this situation is exactly what we described in the $n=2$ and $n=3$ cases---in $n=2$, $p_{H, 1} > 9$ $\forall D$ (see Figure~\ref{n2n3}a,b), while for $n=3$, $p_{H, 1} < 0$ for $D < 7$ and $p_{H, 1} > 0$ for $D \geqslant 8$ (see Figure~\ref{n2n3}c,d). From~Figure~\ref{genn}d one can see that $D_{cr}$ never crosses $n=4$, and analytics can provide us with an exact limit:

$$
\lim\limits_{D\to\infty} n_{cr} = \dac{11 - \sqrt{13}}{4} \approx 3.65.
$$

Outside of these $n=2, 3$ cases, $p_2 < 0$ everywhere else, as~additionally illustrated on panel~(e).

Finally, $p_3$ has a small area of positive values but with no integer solutions (see panel~(f)) (some zeroth that correspond to trivial solutions described in $n=2, 3$ cases), such that we have $p_3 < 0$ apart from the just mentioned cases, as additionally illustrated on panel~(g).

\section{Discussion and~Conclusions}
\textls[10]{Thus, apart from the described $n=2, 3$ cases, general $n \geqslant 4$ Lovelock gravity does not admit Kasner solutions that resemble realistic compactification. This result is somewhat unexpected. Indeed, the~results we obtained so far for the cosmological dynamics in $n=2$ and $n=3$ Lovelock gravity,
demonstrate rise of complexity both for abundance and emergence of regimes---$\Lambda$-term EGB models~\cite{my16b, my17a} demonstrate more different regimes than vacuum EGB models~\cite{my16a}, and vacuum cubic Lovelock models~\mbox{\cite{cubL1, cubL2}} show more regimes than vacuum EGB models. Following this logic, one would expect more different regimes to emerge in $n=4$ Lovelock cosmological models, but apparently, in~this case, we would have even less than in $n=2, 3$ cases---at least, for~power-law solutions. The reasons behind this are not quite clear, so we would need to investigate the $n=4$ case to uncover them.}

\begin{figure}[H]
\centering
\includegraphics[width=1.0\textwidth, angle=0]{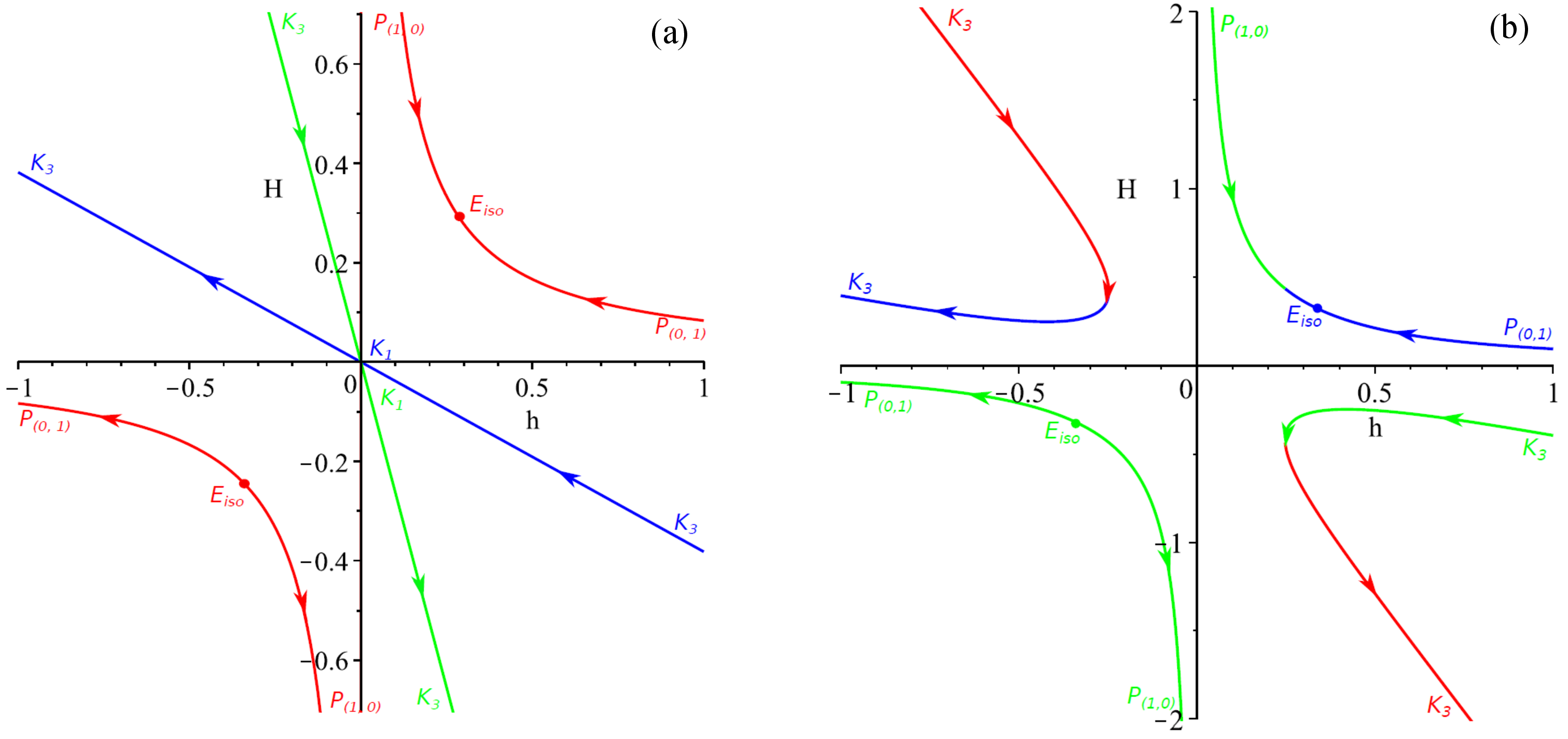}
\caption{$K_3 \to K_3$
 transition lack in EGB $D=3$ the vacuum case (panel (\textbf{a})) and its presence in the EGB $D=3$ $\Lambda$-term case (panel (\textbf{b}))
(see text for more details).}\label{fig3}
\end{figure}

Among the regimes reported in previous studies of particular interest is $K_{2n-1} \to K_{2n-1}$---a transition from one high-curvature Kasner regime to another. We observed this behavior for $\Lambda$-term EGB gravity in $D\geqslant 3$~\cite{my17a}, but it is absent in vacuum EGB gravity~\cite{my16a}. The reason for this is quite interesting. Formally, both regimes ($K_3$ as a past and as a future asymptote) are present and stable in $D \geqslant 3$ EGB cosmology, but a smooth transition between them exists only in the $\Lambda$-term case because, in this case, there is no $K_1$ regime---the low-energy (Einstein-Hilbert) Kasner regime. In~GR, the Kasner solution exists only in a vacuum and is absent in the presence of the $\Lambda$ term, while in EGB, the situation is different. This situation is illustrated in Figure~\ref{fig3}---on panel (a), we present the situation for the $D=3$ vacuum EGB case, and one can see that $K_1$ works as a past asymptote for one of $K_3$ and as a future asymptote for another. On the other hand, in the $D=3$ $\Lambda$-term EGB case, as presented in Figure~\ref{fig3}b, there is no $K_1$, so a smooth transition between one $K_3$ and another is possible.

But what is more, since, for $n \geqslant 4$, all resulting three-dimensional Kasner exponents are negative, no Kasner solutions with realistic compactification exist as a past asymptote, which could hinder the possibility of building realistic compactification: for realistic compactification, we need not only a realistic late-time regime, but also a past regime and a smooth transition between the two. 
Typical realistic compactification in EGB gravity (see~\cite{my24}) is $K_3 \to E_{3+D}$, where $K_3$ is the GB Kasner solution and $E_{3+D}$ is the exponential solution---both with expanding three dimensions and contracting extra dimensions. But our analysis suggests that no such Kasner solutions with expanding three dimensions exist in $n \geqslant 4$ Lovelock gravity---meaning that no such realistic compactification transition exists.

We would expect that the results affect vacuum models more than those with the $\Lambda$ term or perfect fluid. For vacuum models, the entire evolution
happens within the same quadrant on the $(H, h)$ plane:  $\{H, h\}=0$ serve as physical singularities (which could be seen from (\ref{constr_gen})), so during the evolution, $\{H, h\}=0$ cannot be crossed, bounding the entire evolution within the quadrant. Then, since, for $n\geqslant 4$, there are no Kasner regimes with expanding three dimensions and contracting extra dimensions, there is no past regime for the compactification scheme, so the entire scheme fails to work. On the other hand, for~$\Lambda$-term regimes, $\{H, h\}=0$ are not singularities anymore, so they could be crossed, and formally, the past Kasner regime could originate from the quadrant with the ``wrong'' sign for $H$ and $h$ (say, both expanding). If so, this could be really interesting to study how an initially non-compactification regime could evolve into realistic compactification.

On the other hand, $K_3 \to E_{3+D}$ is not the only transition that leads to compactification, but~the most natural, and~its absence hinders compactification in Lovelock gravity. Another possibility for the past asymptotic regime is the power-law regime with one of the Kasner exponents equal to zero (labeled as ``static'' solutions in cases studies), but~it seems that these regimes are not quite stable (for instance, in~cases with the $\Lambda$ term~\cite{my16b, my17a}, it seems that the sign of the $\Lambda$ term affects their asymptotic value $0 \pm 0$, while formally, the $\Lambda$ term should not affect power-law regimes) and are affected by lower-order Lovelock contributions.
Either way, this situation needs to be considered in more detail in papers to follow, as the presence of lower-order (compared to $n=4$) Lovelock contributions could change past or future asymptotes.

In summary, the results reported in the current manuscript are not only interesting by themselves, but they also open a window of possibilities for further research.
Indeed, the absence of Kasner regimes with expanding three dimensions has the potential, if not to prohibit, then to definitely hinder the possibility of describing successful compactification in Lovelock gravity. Lovelock gravity, in~turn, is one of the candidates for the gravitational counterpart of M/string theories, leading
to potential hindering of compactification in these theories, which, in~turn, questions their viability. Of course, our results are obtained for the simplest case with just two subspaces, and both of them are spatially flat, while realistic M/string theory realizations have quite a complicated topology of extra dimensions, but still, such effects in the simplest case could be indicative of potential problems in realistic scenarios and definitely require additional investigation.

\vspace{6pt}
\funding{This research received no external~funding.}

\dataavailability{No new data were created or analyzed in this study}

\conflictsofinterest{The author declares no conflict of~interest.}

\begin{adjustwidth}{-\extralength}{0cm}
\reftitle{References}

\PublishersNote{}
\end{adjustwidth}

\end{document}